\documentstyle[11pt,equations,doublespace,cite,psfig]{article}
\newcommand{\smallz}{{\scriptscriptstyle Z}} 
\newcommand{\smallw}{{\scriptscriptstyle W}} %
\newcommand{\smallh}{{\scriptscriptstyle H}} %
\newcommand{\mz}{m_\smallz}
\newcommand{\mw}{m_\smallw}
\newcommand{\mh}{m_\smallh}
\newcommand{\mt}{m_t}
\newcommand{\dr}{\mbox{$ \delta \rho$}}
\newcommand{\dro}[1]{\mbox{$ \delta \rho^{{\scriptscriptstyle
(#1)}}$}}
\newcommand{\azz}{\mbox{$ A_{\smallz \smallz} $}}
\newcommand{\Azz}[1]{\mbox{$ A_{\smallz \smallz}^{{\scriptscriptstyle
(#1)}}$}}
\newcommand{\aww}{\mbox{$ A_{\smallw \smallw} $}}
\newcommand{\Aww}[1]{\mbox{$ A_{\smallw \smallw}^{{\scriptscriptstyle
(#1)}}$}}
\newcommand{\gmud}{\mbox{$ G_\mu^2 m_t^2 \mz^2 $}}
\newcommand{\gmuq}{\mbox{$ G_\mu^2 m_t^4 $}}
\newcommand{\selfs}{self-energies}
\newcommand{\self}{self-energy}

\newcommand{\ew}{electroweak}
\newcommand{\Dq}{\frac{\partial}{\partial q^2}}
\newcommand{\dq}{\frac{\partial}{\partial q_\mu}}
\newcommand{\dg}{\frac{\delta g^{{\scriptscriptstyle (1)}}}{g}}
\newcommand{\dy}{\frac{\delta Y^{{\scriptscriptstyle (1)}}}{\mz^2}}
\newcommand{\Vwu}[1]{\mbox{$V^{{\scriptscriptstyle (#1)}}_\smallw$}}
\newcommand{\Bwu}[1]{\mbox{$B^{{\scriptscriptstyle (#1)}}_\smallw$}}
\newcommand{\Vzu}[1]{\mbox{$V^{{\scriptscriptstyle (#1)}}_\smallz$}}
\newcommand{\Bzu}[1]{\mbox{$B^{{\scriptscriptstyle (#1)}}_\smallz$}}
\newcommand{\ik}{\int \frac{d^n k}{(2 \pi)^n\, \mu^{n-4}}}
\newcommand{\tint}{\int d^n y \,e^{-iq \cdot y} \int d^n x_1 \,e^{ik
\cdot x_1}
                  \int d^n x_2 \, e^{-ik \cdot x_2}}
\newcommand{\dint}{\int d^n y \,e^{-i q \cdot y}  \int d^n x \,e^{ik
\cdot x}}
\newcommand{\uint}{\int d^n x \,e^{ik \cdot x}}
\newcommand{\qua}[4]{\langle #1 (y)\, #2 (0) \, #3 (x_1) \,
                     #4 (x_2) \rangle}
\newcommand{\quaesteso}[4]{\langle 0 \, | \,T^* \left[ #1 (y)\, #2
(0) \, #3 (x_1) \,
                     #4 (x_2) \right]\,  |\, 0 \rangle}
\newcommand{\tre}[3]{\langle #1 (y)\, #2 (x) \, #3 (0)  \rangle}
\newcommand{\due}[2]{\langle #1 (x) \, #2 (0)  \rangle}
\newcommand{\equ}[1]{eq.~(\ref{#1})}
\newcommand{\eqs}[1]{eqs.~(\ref{#1})}
\newcommand{\Eqs}[2]{eqs.~(\ref{#1}) and (\ref{#2})}
\newcommand{\efe}[1]{ref.\cite{#1}}
\newcommand{\efs}[2]{refs.\cite{#1,#2}}
\newcommand{\be}{\begin{equation}}
\newcommand{\ee}{\end{equation}}
\newcommand{\een}{\end{subequations}}
\newcommand{\ben}{\begin{subequations}}
\newcommand{\beq}{\begin{eqalignno}}
\newcommand{\eeq}{\end{eqalignno}}
\newcommand{\beqs}{\begin{eqalignno*}}
\newcommand{\eeqs}{\end{eqalignno*}}
\newcommand{\bea}{\begin{eqnarray}}
\newcommand{\eea}{\end{eqnarray}}

\renewcommand{\thefootnote}{\fnsymbol{footnote} }
\begin{document}
\begin{titlepage}
\begin{flushright}
	\small
 	CERN-TH.7180/94\\
        DFPD 94/TH/12\phantom{94}\\
        NYU-Th-94/02/01
\end{flushright}

\begin{center}
{\Large Two-loop next-to-leading $\mt$ corrections to the $\rho$
parameter}
 \\
\vspace{1.5cm}
Giuseppe Degrassi\\
{\em Dipartimento di Fisica, Universit\`a  di Padova,
Sezione INFN di Padova \\
Via Marzolo 8, 35131 Padova, Italy} \\
Sergio Fanchiotti \\
{\em Theory Division, CERN, CH-1211 Geneva 23, Switzerland}\\
Paolo Gambino\\
{\em Department of Physics, New York University, New York, NY 10003,
USA}
\end{center}
\vspace{2cm}
\begin{center}
{\bf Abstract}
\end{center}
\vspace{0.2cm}
The $O(\gmuq)$ correction to the $\rho$ parameter is computed within
the
Standard Model using the current algebra formulation of radiative
corrections.
This approach is proved to be equivalent to the effective Lagrangian
method
proposed by Barbieri {\em et al.} Using the same framework, the
$O(\gmud)$
correction to the ratio of neutral-to-charged current amplitudes is
analysed in an $SU(2)$ model. The $O(\gmud)$ contribution is shown to
be
numerically comparable to the leading $O(\gmuq)$ term for realistic
values
of the top mass. The resummation of higher-order effects is
discussed.\par
\vspace{0.2cm}\nopagebreak
{\small \noindent CERN-TH.7180/94\\ February~1994}
\end{titlepage}
\setcounter{footnote}{0}
\renewcommand{\thefootnote}{\arabic{footnote}}

%
%
\section{Introduction}
The key role played by a heavy top quark in the analysis of \ew\ data
has
by now been emphasized for more than fifteen years \cite{b1}.
One-loop
virtual top effects are present in the \selfs\ of the $W$ and $Z$
vector
bosons \cite{b1,b2} and in the $Zb \bar{b}$ vertex \cite{b3}.
Self-energy
contributions affect important \ew\ quantities like the $\rho$
parameter,
which measures the relative strength, at low momentum transfer,
between the
neutral- and charged-current interactions
\be
\rho = 1 + \dr \label{e1i}
\ee
and $\Delta r$ \cite{b4}, the correction relevant to the $\mw$--$\mz$
interdependence.

As is  well known, at the one-loop level the top contribution to \dr\
is given by  \cite{b1}
\ben \label{e2i} \beq
\dr =& N_c x_t \\ \label{e2ai}
x_t =& \frac{G_\mu \mt^2}{8\, \pi^2\, \sqrt2} \label{e2bi}
\eeq \een
where $N_c$ is the colour factor,
while the leading top behaviour in $\Delta r$ can be described by
\cite{b2,b4}
\be
\Delta r \simeq {c^2 \over s^2} \dr \,, \label{e3i}
\ee
where $s$ is an abbreviation for $\sin \theta_W \equiv g^\prime /
\sqrt{g^{\prime 2} + g^2}$,  $c^2= 1-s^2$, and $g$ and $g^\prime$ are
the
coupling constants of $SU(2)$ and $U(1)$ respectively.

Since \dr\ increases rapidly with $\mt$, it is possible to get an
upper
bound on the top mass using \ew\ precision measurements. A recent
analysis
of the available precision data yields $\mt= 164^{+16+18}_{-17-21}$
GeV
with $\mh= 300^{+700}_{-240}$ GeV \cite{b5}. This bound applies
within
the Standard Model (SM) or assuming that virtual effects, due to
other
kinds of undiscovered physics, contribute very little to \ew\
quantities.

With the increasing evidence that the top is heavy, attention has
been paid to
two-loop corrections involving this quark. Following the work of van
der
Bij and Hoogeveen on the two-loop leading top effect to the $\rho$
parameter,
namely the $O(\gmuq)$ contribution, QCD corrections to the one-loop
$\mt^2$
term have been computed both for the \selfs\ of the vector bosons
\cite{b7}
and for the $Z \rightarrow b \bar{b}$ decay width \cite{b8}.
Concerning the
latter,
the \ew\ two-loop $\mt^4$ corrections are now also available
\cite{b9,b10,b11}.

The first calculation \cite{b6} of the one-particle irreducible (1PI)
$O(\gmuq)$ contribution to the $\rho$ parameter was performed a few
years ago
in the approximation of a small Higgs mass, $\mh \ll \mt$. Recently
it was
extended by Barbieri {\em et al.} to the case of arbitrary $\mh$.
Subsequently,
a simple explicit analytic expression for the Higgs-mass correction
was derived
\cite{b11}.
An interesting feature of the leading $\mt$ corrections, first noted
in
\efe{b11bis} and then developed to two-loops in \efe{b9}, is that
these terms
have very little to do with the gauge sector of the SM.
Indeed the calculations presented in \efe{b9}\ and \efe{b11}\ were
performed
using
a Yukawa Lagrangian involving a Higgs doublet plus the top and bottom
fields.
The connection between physical quantities and the renormalization
constants
of the Yukawa Lagrangian was obtained by exploiting the Ward
identities
of the SM Lagrangian with the gauge interaction switched off. The use
of
this ``gaugeless'' limit greatly simplifies the two-loop calculation,
making it quite manageable.

The aim of this paper is twofold. First, we derive explicitly the
connection
between the SM and the effective Yukawa Lagrangian used in \efe{b9}.
We
consider the SM diagrams contributing to \dr\ to leading order in
$\mt$
and we evaluate them employing techniques used in the current-algebra
formulation of radiative corrections \cite{b12}. This framework
allows
us to easily enforce the relevant Ward identities, leading us to a
result
that differs from the one presented in \efs{b9}{b11} by subleading
contributions
$O(\gmud)$. These terms are clearly due to the gauge sector of the
theory,
as can be  seen by noticing that the $O(\gmuq)$ contribution
corresponds
to effects proportional to $g_t^4$, where $g_t$ is the top Yukawa
coupling,
while a $O(\gmud)$ term scales as $g_t^2 g^2$, with $g$ the gauge
coupling.
As a second result we perform a complete calculation (reducible plus
irreducible parts) through $O(\gmud)$ of the $\rho$ parameter in an
$SU(2)$ theory. Although this result has no direct implication
in the analysis of \ew\ precision experiments, it is meant to provide
an
estimate of the size of subleading two-loop contributions involving
the top.

The plan of this paper is as follows. Section 2 presents the
derivation
of the 1PI $O(\gmuq)$ term in the $\rho$ parameter within the SM,
using
the formalism of current correlation functions and their associated
current algebra. In section 3 we derive the 1PI $\mt$ \selfs\
contribution
to \dr\ in an  $SU(2)$ theory. Section 4 is devoted to the
calculation of the
$\rho$ parameter in $SU(2)$. Finally, in section 5 we discuss the
resummation
of higher-order effects and draw our conclusions.
\section{One-particle irreducible O(\gmuq) correction  to \dr\ in
\newline
        $SU(2) \times U(1)$}
In this section we derive the 1PI\footnote{As the mass counterterm
diagrams
are needed to cancel divergences in the irreducible two-loop
corrections,
we include them in the latter.}
two-loop $m_t^4$ contribution
to \dr\ in the SM, keeping the Higgs mass arbitrary. As far as only
leading terms are concerned, we can identify \dr\ with $\Delta$,
where
\be
\Delta \equiv \frac{\aww (0)}{\mw^2} - \frac{\azz (0)}{\mz^2} .
\label{e1.0}
\ee
In \equ{e1.0} $A(q^2)$ represents the transverse part of the \self,
i.e.\ the
cofactor
of $g^{\mu \nu}$ in the \self\ tensor $\Pi^{\mu \nu} (q^2)$.

Just by power-counting inspection it is easy to realize that the only
diagrams that can contribute to the leading $m_t^4$ term are those
containing,
besides top and bottom, the physical and unphysical scalars. A
suitable
choice of the tadpole counterterm allows one to neglect the tadpole
diagrams in the calculation \cite{b13}. Enforcing it, the only
topologies in the \selfs\
contributing to the leading term are those depicted in fig.~1. In the
figure, wavy lines represent vector bosons, dashed lines scalars
(physical or unphysical) while solid lines are fermions.
\par In order to fix our notation we write the part of the
SM Lagrangian density that describes the interaction of the $W,\,Z$
and  scalars with fermions as \cite{b12}
\be
{\cal L}_{int}  = - \frac{g}{\sqrt{2}}(W^{\dagger}_\mu J_W^\mu +
{\rm h.c.})
        -  \frac{g}{c}Z_\mu J^\mu _Z
        - \frac{g}{2\, \mw} \left[ \Phi_1 S_1 + \Phi_2 S_2
        + \sqrt2 \,( \Phi^\dagger S + {\rm h.c.}) \right] \label{e1}
\ee
where  $\mw$ stands for the mass of the $W$ boson,
$J_Z^\mu$ and $J_W^\mu$ are the
fermionic currents coupled to $Z$ and $W$ respectively,  $W^\dagger$
is the
field that creates a $W^+$ meson, $\Phi_1$ is the physical Higgs
boson,
$\Phi_2$ and $\Phi$ the unphysical
counterparts associated with the $Z$ and $W$ and
\ben \label{e2} \beq
S_1 =&  \bar{\psi}\, m^0\, \psi     \label{e2a0} \\
S_2 =& 2 \,\partial_\mu J^\mu_Z = -i \bar{\psi}\, m^0\, C_3 \gamma_5
\psi
                   \label{e2a} \\
S =& -i \, \partial_\mu J^\mu_W = \bar{\psi} \Gamma \psi. \label{e2b}
\eeq
\een
In \eqs{e2}, $\psi$ represents the column vector $\psi \equiv
(t,b)^T, \:
m^0, \: C_3$ and $\Gamma$ are the $2 \times 2$ matrices
\ben \label{e2cde} \beq
m^0 =& \left(  \begin{array}{cc} m_t^0 & 0 \\
                                  0 & 0 \end{array} \right)
\label{e2c} \\
C_3 =& \left(  \begin{array}{cr} 1 & 0 \\
                                 0 & -1 \end{array} \right)
\label{e2d} \\
\Gamma =& \left(  \begin{array}{cc} 0 & 0 \\
                   -m_t^0 \, a_+ & 0 \end{array} \right) ,
\label{e2e}
\eeq \een
$a_+ \equiv \frac{1+\gamma_5}{2}$, and the superscript
$0$ on $m_t$ refers to the bare mass.
As is evident from \Eqs{e2c}{e2e} we are considering only the third
generation and taking the bottom quark as massless.

We begin by studying \azz (0). Using current correlation functions we
can
combine the amplitudes of fig.~1a,b, where the continuous line
represents a top
and the dashed one a Higgs or $\Phi_2$, into the expression
\beq
\lefteqn{\Pi_{(a,b)}^{\mu \nu}(q^2) = \sum_{j=1}^2  \frac{g^4}{4
\,c^2\,\mw^2}
\frac{1}{2} \ik \frac1{k^2 - m_j^2}} \nonumber \\
& \times  \tint \quaesteso{J_Z^\mu}{J_Z^\nu}{S_j}{S_j} ,\label{e3}
\eeq
where $n$ is the space-time dimension, $\mu$ is the 't Hooft mass
scale, $T^*$ is the covariant time-ordered product and $m_1 \equiv
\mh$
and $m_2 \equiv \mz$. In the case of the unphysical scalar,
eq.(\ref{e3})
is valid in the 't Hooft--Feynman gauge.  As the leading $m_t^4$ term
does
not depend upon the gauge sector we have chosen this gauge  for
practical
convenience.

In order to trigger  Ward identities we contract $\Pi_{(a,b)}^{\mu
\nu}(q^2)$
with $q^\mu\, q^\nu$. Contraction of a  current operator with its
four-momentum gives rise to a term involving the divergence of a
current plus
an equal time commutator that reduces the number of operators inside
the
time-ordered product by one unit. Noticing that
\be
\Dq \left\{ q_\mu q_\nu \Pi^{\mu \nu} (q^2) \right\}_{q^2=0} = A(0),
\label{e4}
\ee
and introducing the short-hand notation $\langle ...\rangle $
for the vacuum expectation value (v.e.v.) of the covariant
time-ordered
product
$\langle 0\, |\, T^* ... |\, 0 \rangle$,
we can write
\beq
\lefteqn{A_{(a,b)} (0) = \frac{g^4}{8\, c^2 \mw^2} \Dq \ik \int d^n y
e^{-iq \cdot y} } \nonumber \\
&\times \left\{ \sum_{j=1}^2 \int d^n x_1 e^{i k \cdot x_1} \int d^n
x_2
e^{-ik \cdot x_2} \frac{\qua{S_2}{S_2}{S_j}{S_j}}{4(k^2 - m_j^2)}
\right. \nonumber\\
&\left. \mbox{}+ i \uint \tre{S_2}{S_1}{S_2} \left[ \frac1{k^2 -
m_1^2} -
\frac1{(k-q)^2 - m_2^2} \right] \right\} _{q^2=0} \nonumber\\
&\mbox{}+ \frac{g^4}{16\, c^2 \mw^2} \ik \left\{ \left(1 - {4 \over
n}
\frac{k^2}{k^2- m_1^2} \right) \uint \frac{\due{S_2}{S_2}}{(k^2 -
m_1^2)^2}
+ (1\leftrightarrow 2)
\right\} . \label{e5}
\eeq
In \equ{e5}\ the notation $ (1 \leftrightarrow 2)$ represents a term
obtained
by the  previous one inside the curly bracket by the substitution
$1 \leftrightarrow 2$.

We now examine the contributions that are described by the topology
shown in fig.~1c. Let us consider the case when one dashed line
represents a $\Phi_1$ and the other one a $\Phi_2$. We write
\beq
\Pi_{(c)}^{\alpha \nu}(q^2) =& - \frac{g^4}{4\, c^2 \mw^2} \ik
    \frac{(2k-q)^\nu}{[k^2-m_1^2][(k-q)^2-m_2^2]} \nonumber \\
 & \times \dint  \tre{J_Z^\alpha}{S_1}{S_2} \label{e6}.
\eeq
Contraction with $q^\alpha$ gives
\beq
\lefteqn{q_\alpha \Pi_{(c)}^{\alpha \nu} (q^2)=  \frac{i g^4}{8\, c^2
\mw^2}
\ik  \frac{(2k-q)^\nu}{[k^2-m_1^2][(k-q)^2-m_2^2]}} \nonumber \\
&  \times \left\{ \dint  \tre{S_2}{S_1}{S_2} \right. \nonumber\\
&\mbox{}+i \left. \int d^n x \left[ e^{i(k-q) \cdot x} \due{S_2}{S_2}
- e^{i k \cdot x} \due{S_1}{S_1} \right] \right\}.
\label{e7}
\eeq
Recalling that
\be
\dq \left(q_\alpha \Pi^{\alpha \nu} (q^2) \right) = \Pi^{\mu
\nu}(q^2) +
q_\alpha \dq \Pi^{\alpha \nu} (q^2) \label{e8}
\ee
it follows that the contribution of fig.~1c to $\Pi^{\mu
\nu}_{ZZ}(0)$ is
obtained by differentiating \equ{e7} with respect to $q^\mu$ and then
setting $q^2=0$. Consider the first term in \equ{e7},
\be
T^\nu = \ik  \frac{(2k-q)^\nu}{[k^2-m_1^2][(k-q)^2-m_2^2]}
\big\{...\big\}
\label{e9}
\ee
where $\big\{...\big\}$
 represents the three-point correlation function times the
appropriate
constants. By Lorentz invariance, $T^\nu$ should have the form
$T^\nu = \alpha (q^2) q^\nu$, 
and therefore
\be
\dq T^\mu = 2\frac{\partial\alpha}{\partial q^2}( q^2)\
 q^\mu q^\nu + \alpha (q^2)\  g^{\mu \nu}. \label{e10}
\ee
The contribution to $A(q^2)$ is then given by $\alpha(q^2)$. It is
easy
to show that
\be
\alpha (0) = \left. \Dq ( q_\nu T^\nu) \right|_{q^2=0}, \label{11}
\ee
or  explicitly
\be
\alpha(0) = \Dq \left. \ik
\left[ \frac1{(k-q)^2- m_2^2} - \frac1{k^2- m_1^2}
           + \frac{m_1^2 - m_2^2}
{(k^2 - m_1^2)[(k-q)^2 - m_2^2]} \right]
\  \big\{...\big\}\right|_{q^2=0}. \label{e13}
\ee

Using \equ{e13}, 
we can write for the transverse part of $\Pi_{(c)}^{\mu \nu}\,$:
\beq
\lefteqn{A_{(c)} (0) = \frac{i g^4}{8\, c^2 \mw^2} \Dq \left\{
    \dint \tre{S_2}{S_1}{S_2}
\right. } \nonumber\\& \times \left.
\ik
\left[\frac1{(k-q)^2- m_2^2} - \frac1{k^2- m_1^2}+
 \frac{ m_1^2 - m_2^2}
{(k^2 - m_1^2)[(k-q)^2 - m_2^2]} \right]
              \right\}_{q^2=0} \nonumber \\
 &\mbox{}- \frac{g^4}{8\, c^2 \mw^2} \ik \left\{ \left(1 - {4 \over
n}
       \frac{k^2}{k^2- m_1^2} \right) \uint
       \frac{\due{S_2}{S_2}}{(k^2 - m_1^2)(k^2- m_2^2)}
+ (1 \leftrightarrow 2)
 \right\} .\nonumber \\
& \  \label{e14}
\eeq
The contribution of $\Phi_1$ and $\Phi_2$ to fig.~1d,e can be
similarly
computed:
\beq
A_{d,e} (0) =& \frac{g^4}{16\, c^2 \mw^2} \ik \left\{ \left(1 - {4
\over n}
       \frac{k^2}{k^2- m_2^2} \right) \uint
       \frac{\due{S_1}{S_1}}{(k^2 - m_1^2)^2}
+ (1\leftrightarrow 2 )
\right\} . \nonumber \\
& \ \label{e15}
\eeq
An analogous  analysis for the $\Phi$ contribution to  $\azz (0)$
gives
\beq
\lefteqn{A_{(a-e)} (0) = \frac{g^4}{8\, c^2 \mw^2} \Dq \left\{ \ik
\frac1{k^2 - \mw^2} \right. } \nonumber \\
& \times \left. \tint \qua{S_2}{S_2}{S^\dagger}{S} \right\}_{q^2=0}.
\label{e16}
\eeq

Summing eqs.~(\ref{e5}), and (\ref{e14})--(\ref{e16}) we get,
for the transverse part of the $Z$ \self\ at $q^2=0$,
\ben \label{e17} \beqs
\lefteqn{\frac{\azz (0)}{\mz^2} =
{g^4\over 8 \mw^4} \Dq \ik \left\{ \tint \right. } \\
&\times \left[ { \qua{S_2}{S_2}{S^\dagger}{S} \over k^2-\mw^2} +
\sum_{j=1}^2 { \qua{S_2}{S_2}{S_j}{S_j} \over 4 (k^2-m_j^2)} \right]
\\
&\left. +i(\mh^2-\mz^2)  \dint { \tre{S_2}{S_1}{S_2} \over
(k^2-\mh^2)((k-q)^2-\mz^2)} \right\}_{q^2=0} \\
&\mbox{}+ {g^4\over 16 \mw^4} \ik
{(\mh^2-\mz^2)^2 \over (k^2-\mh^2)^2(k^2-\mz^2)^2}
\\& \times
\left\{ \left( 1- {4 \over n}{k^2 \over k^2 - m_2^2} \right) \uint
\due{S_1}{S_1} + (1\leftrightarrow 2)
\right\} \yesnumber \label{e17a} .
\eeqs

The discussion of the $W$ \self\ can be performed on the same footing
as
the $Z$ case. Here we present only the result
\beqs
\lefteqn{\frac{\aww (0)}{\mw^2} =
{ g^4\over 8 \mw^4 } \Dq \ik \left\{ \tint \right. } \\
&\times\left[ { \langle S^\dagger (y)S(0)\left(S^\dagger (x_1)S(x_2)+
{\rm h.c.}
\right) \rangle \over(k^2-\mw^2)} +
\sum_{j=1}^2 { \qua{S^\dagger}{S}{S_j}{S_j} \over 2(k^2-m_j^2)}
\right] \\
&\mbox{}+ \dint \left[  i(\mh^2-\mw^2) {\langle \left(
S^\dagger(y)S(0)+ {\rm h.c.}\right)S_1(x)\rangle \over
(k^2-\mh^2)((k-q)^2-\mw^2)} \right.\\
&\hphantom{\dint [} \mbox{}+ \left. \left. (\mz^2-\mw^2) {\langle
\left( S^\dagger(y)S(0) -{\rm h.c.} \right) S_2(x)\rangle
\over (k^2-\mz^2)((k-q)^2-\mw^2)}\right] \right\}_{q^2=0}\\
&\mbox{}+ {g^4\over 16 \mw^4 } \ik \left\{
{(m_1^2-\mw^2)^2 \over (k^2-m_1^2)^2(k^2-\mw^2)^2}
\left[ \left( 1- {4 \over n}{k^2 \over k^2 - \mw^2} \right) \uint
\due{S_1}{S_1} \right.\right.\\
& \hphantom{\dint [}
+ \left.\left. 2 \left( 1- {4 \over n}{k^2 \over k^2 - m_1^2} \right)
\uint
\due{S^\dagger}{S} \right]+  (1\leftrightarrow 2) \right\}
\yesnumber \label{e17b} .
\eeqs
\een

Before discussing the diagrams involving the counterterms,
it is interesting to consider
the limit $g,\,g^\prime \rightarrow 0$ in \eqs{e17} and the
connection with
the result
presented in \efe{b9}. The authors of \efe{b9}\ observed that, for
what concerns
the leading one- and two-loop $\mt$ effects, there is a Ward identity
that
relates, at $q^2=0$,  the $W$ and $Z$ \selfs\ to the ones of the
unphysical
counterparts. Explicitly
\ben \label{e18} \beq
\frac{\aww (0)}{\mw^2} =& - \left. \Dq \Pi_{\Phi \Phi} (q^2)
\right|_{q^2=0}
                        \label{e18a} \\
\frac{\azz (0)}{\mz^2} =& - \left. \Dq \Pi_{\Phi_2 \Phi_2} (q^2)
                       \right|_{q^2=0} .               \label{e18b}
\eeq \een
As noted in \efe{b9}, \eqs{e18} have the advantage that the r.h.s.\
can be
evaluated by switching off the gauge interaction in the SM
Lagrangian, namely
using only the Yukawa and Higgs part of it. In fig.~2 we show the
diagrams
belonging to the scalar \selfs\ that contribute to the leading
$m_t^4$ term.

In order to understand the connection between \eqs{e17} and the
r.h.s.\
of \eqs{e18}
consider for example the four-point correlation functions in
\equ{e17a}.
In the limit $g,\,g^\prime \rightarrow 0$ or $\mw,\, \mz \rightarrow
0$,
their contribution to \dr\ is proportional to $g_t^4$. In fact every
$S$ operator contains an $\mt=g_t v$ factor, where $v$ is the
v.e.v., while the coefficient in front is proportional
to $1/v^4$. In this limit the terms involving four-point functions
represent exactly the contribution to $\left. \Dq \Pi_{\Phi_2 \Phi_2}
(q^2)
\right|_{q^2=0}$ of
the diagrams shown in
figs.~2a and 2b. A similar connection can be established
between the three- and two-point correlation functions in
\equ{e17a} and the
diagrams 2c and 2d respectively. We notice that figs.~2c and 2d
involve trilinear scalar couplings proportional to $\mh^2$. Although
no such   coupling is present in the diagrams of fig.~1, we recover
these terms using \equ{e13} and simple algebraic manipulations.
Equation (\ref{e17b}) shows an analogous connection between its
various
contributions and $\left. \Dq \Pi_{\Phi \Phi} (q^2) \right|_{q^2=0}$
computed
in the Yukawa theory. We observe that, in obtaining \eqs{e17}, we
took
advantage of
several commutation relations that were derived using an
anticommuting
$\gamma_5$.

To complete the calculation of the leading $\mt^4$ term, we have to
include
the counterterm diagrams. The counterterms that can contribute to the
order
we are interested in are the mass counterterm for the top, Higgs and
unphysical scalars\footnote{The mass counterterm for the unphysical
scalars is related to the tadpole counterterm and therefore
determined
from the tadpole diagrams \cite{b13}.}. However, it is easy to see
that the
diagrams containing the counterterm associated to the scalars do not
give $\mt^4$
terms in \dr\ because of the cancellation of the leading part between
the
two \selfs. We are left with the mass counterterm of the top quark,
which we define on-shell. As we are studying the  $\mt^4$ term, it is
sufficient to include in the top counterterm only diagrams involving
scalars.
In the 't Hooft--Feynman gauge we write the contribution
to $\Delta$ arising from the diagrams containing the top counterterm
as
\beq
\Delta_{ct}=& N_c x_t^2 \left\{ 3 \left[ \frac1{\epsilon} + 2\,{\cal
C}
        - 2 \ln \frac{\mt^2}{\mu^2} \right] + \frac92 -ht
\frac{6-ht}2
         \ln ht - ht \right. \nonumber  \\
& \left.   + \frac{ht-4}2 \sqrt{ht}\, g(ht) + zt \left( \frac{zt}2-1
\right)
  \ln zt - zt +{ zt^{3/2} \over 2} g(zt) - 2 N(\sqrt{wt}) \right\}\,,
  \label{e19}
\eeq
where $n= 4- 2 \epsilon$ and ${\cal C}= -\gamma_E +\ln 4 \pi$,
$ht= \mh^2 / \mt^2, \: zt=\mz^2 / \mt^2,\: wt=\mw^2 / \mt^2$,
the function $N$ is defined in Appendix A of \efe{b14}, and
\beq
g(x) =&  \left\{
          \begin{array}{lr}
           \sqrt{4-x}\,\left
	      (\pi - 2 \arcsin{\sqrt{x/4}}
	               \right) & 			0 < x \leq 4 \\
	   {}\\
	   2 \sqrt{x/4-1}\,\ln\left(
	   	\frac{1-\sqrt{1-4/x}}{1+\sqrt{1-4/x}}
		              \right) & 		x > 4 \,.
	  \end{array}
	\right.  \label{e19a}
\eeq

The 1PI $\mt^4$ contribution to \dr\ is obtained by working out
explicitly
the correlation functions in \eqs{e17} and then evaluating the
resulting
integrals and \equ{e19} in the limit $\mw=\mz=0$. In the actual
computation
we have used an anticommuting $\gamma_5$ and
some recent contributions on two-loop integrals \cite{b15}. We
find perfect agreement with the analytic results presented in
\efe{b11}.

\section{Self-energy irreducible contribution to \dr\ in SU(2)}
In the previous section we applied current algebra techniques to
simplify
diagrams contributing to the leading $\mt^4$ term: in this way we
obtained
expressions (\eqs{e17}) containing not only the leading term, but
also
subleading contributions. Obviously, there are other 1PI diagrams
that
can contribute to subleading order.

However, when one tries to go beyond the leading $\mt$ term in \dr,
the calculation of  $\Delta$ is no longer sufficient and one has
to investigate  the ratio of neutral-to-charged-current scattering at
zero
momentum transfer. As a consequence, vertex and box diagrams should
also be
taken
into account and a complete two-loop calculation becomes a very
difficult
task.

Here we present a first study of the next-to-leading-order two-loop
$\mt$ corrections to \dr\ in an $SU(2)$ model obtained by setting
$g^\prime =0$
in the SM.   This amounts
to computing the 1PI and reducible \selfs, vertex and box diagrams
present in
neutral and charged scattering processes to $O(\gmud)$. We recall
that in the
$SU(2)$ model the bare masses of the charged and neutral vector
bosons
are equal, there is no weak mixing angle, and electromagnetic
interactions
are not present. The breaking of the symmetry by a non-degenerate
$(t,b)$
doublet is responsible for the splitting in the values of the
physical
masses of the vector bosons.

In this section we study the 1PI two-loop \selfs \footnote{The
treatment
of the counterterm diagrams and tadpoles is analogous to
what was done in section 2.};
in particular we concentrate on\footnote{We indicate the charged and
neutral
vector bosons of $SU(2)$ as
$W$ and $Z$, respectively. Their \selfs\ are correspondingly labelled
$\aww$ and $\azz$.}
$\Delta_2 \equiv [\aww(0)- \azz(0)]/ \mz^2$.
Using the results of \efe{b15}, the
$\mt$ contribution to it can be actually computed without any
approximation.
However, we report only the first  correction, $O(zt)$, to the
leading term
in order to be consistent with the accuracy we compute the
remaining parts contributing to \dr. We begin by analysing the
contribution
of the scalars to the
\selfs. Beside the diagrams of fig.~1, to $O(\gmud)$ there are other
graphs
containing the Higgs boson that have to be considered. They are shown
in
fig.~3.
To evaluate these new contributions we use the same
techniques as employed in the previous section, namely we  express
groups
of Feynman diagrams in terms of current correlation functions and
then
contract the currents inside with their four-momenta to enforce Ward
identities. In order to use the results of the previous section we
also employ in the calculation of the subleadings the 't
Hooft--Feynman
gauge.
We combine \eqs{e17} and (\ref{e19}), evaluated in the limit
$\mw=\mz$,  $c=1$, with the diagrams of fig.~3 , and obtain
\ben \label{e20} \beqs
\Delta_2^s &= N_c x_t^2 \left\{ 25 -4ht -8zt +{4 zt \over ht} + \pi^2
\left( \frac12
- \frac1{ht} +\frac{2 zt}3 - {2 zt \over 3 ht^2} + {zt \over ht}
\right)
+ \frac{(ht-4) \sqrt{ht} \,g(ht)}2 \right. \\
&\mbox{}+ {(ht-1)^2 (6ht- 3ht^2 +4 zt +2ht zt) Li_2 (1-ht) \over
ht^2}
- \left( 6+6 ht - \frac{ht^2}2 - 2 zt + {4 zt \over ht} \right) \ln
(ht) \\
& \left. -7 zt \ln (zt) +\left( -15 +9 ht - {3 ht^2 \over 2} - 2 zt -
{8 zt \over ht} + ht zt \right) \phi \left( \frac{ht}4 \right)
\right\} ,
\yesnumber  \label{e20a}
\eeqs
for the two-loop contribution to $\Delta_2$ involving the top and the
scalars to $O(zt)$, in the region $ht \gg zt$  ,
while for $ht \ll zt \ll 1$:
\beq
\Delta_2^s = N_c x_t^2 \left\{ 19 - 2 \pi^2 -4 \pi \sqrt{ht} - 16 zt
+
          {2 \pi^2 \over 3} zt - zt \ln zt \right\} \,, \label{e20b}
\eeq \een
In \equ{e20a}
\ben \label{e21} \beq
Li_2 (x) =& - \int_0^x dt {\ln (1-t) \over t}     ,   \label{e21b}
\eeq
and
\be
\phi(z) =
       \begin{cases}
       4 \sqrt{{z \over 1-z}} Cl_2 ( 2 \arcsin \sqrt z ) & $ 0 < z
\leq 1$\\
       { 1 \over \lambda} \left[ - 4 Li_2 ({1-\lambda \over 2}) +
       2 \ln^2 ({1-\lambda \over 2}) - \ln^2 (4z) +\pi^2/3 \right]
       & $z >1 $\,,
       \end{cases}
       \label{e21c}
\ee
where $Cl_2(x)= {\rm Im} \,Li_2 (e^{ix})$ is the Clausen function
and
\be
\lambda = \sqrt{1 - {1 \over z}}. \label{e21d}
\ee
\een
We notice that when $zt \rightarrow 0$ in \equ{e20a}, $\Delta_2^s$
becomes
identical to eq.~(12) of \efe{b11}.

The remaining diagrams belonging to the 1PI \selfs\ are obtained by
replacing all dashed lines in fig.\ 1 by wavy ones, namely
substituting the
unphysical scalars by their associated vector bosons. We consider
this
contribution together with the part of the counterterm diagrams not
included
in \eqs{e20} as there is a partial cancellation in the ultraviolet
poles.
Concerning the counterterms, we showed in section 2 that in the SM
the
diagrams containing the mass counterterm of the scalars
do not contribute
to the leading $\mt^4$ term. In the restricted $SU(2)$ model we
are now considering, the bare masses, $\mw^0$ and $\mz^0$, are equal
and only
one mass counterterm is available in the gauge-boson sector.
Consequently,
vector boson and Higgs mass counterterm graphs cancel out exactly in
$\Delta_2$
and we are left only with diagrams involving the top-mass
counterterm.
Equations (\ref{e20}) include diagrams containing the part of the
top-mass
counterterm due to the scalars; to $O(\gmud)$ one has to consider in
the counterterm also the contribution of the vector bosons. The
diagrams
involving this part of the top counterterm together with the diagrams
of
fig.~1, assuming the internal dashed lines to represent vector
bosons, give
to $O(zt)$
\be
\Delta_2^{v.b.} = N_c x_t^2 \left\{ 16 zt \left( {1 \over \epsilon} -
2 \ln \left( {\mt^2 \over \mu^2} \right)
+ 2\, {\cal C} \right) + 56 zt + 6 zt \ln zt - 3 \pi^2 zt
                        \right\} .       \label{e23}
\ee
We notice that $\Delta_2^{v.b.}$ contains diagrams involving
vector bosons coupled to a fermionic triangle. However, the group
$SU(2)$
is automatically-anomaly free and therefore these graphs
are not affected by the anomaly problem.

The calculation of the two-loop $\mt$ part in $\Delta_2$ has given a
somewhat surprising result. The answer is not finite but a $1 /
\epsilon$ pole has appeared (cf.\ \equ{e23}). This pole is the
confirmation
that, beyond the leading term, $\Delta_2$ cannot be identified  with
\dr,
but the ratio of the complete neutral-to-charged amplitudes has to be
considered. As
we will see in the following section, the vertex contributions
contain
a similar pole (with opposite sign!) and, as expected, \dr\ comes out
finite.

\section{The $\rho$ parameter to $O(\gmud)$ in $SU(2)$}
In this section we compute the ratio of neutral-to-charged-current
amplitudes
to two-loop $O(\gmud)$ contributions in an $SU(2)$ model. We begin by
recalling that in $SU(2)$, concerning the gauge sector, there are
only
two bare parameters, $g_0$, the coupling constant, and $v_0$, the
v.e.v.
Therefore two renormalization conditions are needed. Our strategy is
to
trade the renormalized parameters $g$ and $v$ for the physical
quantities
$G_\mu$ and $\mz$, where $\mz$ is the on-shell mass of the neutral
vector
boson. The simplest way to achieve this is to take
$\delta \mz^2={\rm Re}\azz(\mz^2)$
and adjust $\delta g$ in such a way that the relation
\be
\frac{G_\mu}{\sqrt{2}}=\frac{g^2}{8\mz^2} \label{e25}
\ee
holds up to second order.

Equation (\ref{e25}) implies that the counterterm $\delta g$ should
cancel the
one and two-loop contributions to muon decay. Defining
\be
\delta Y = {\rm Re} \left[\aww(0)-\azz(\mz^2)\right] \label{e26}
\ee
and referring to fig.~4 for the one-loop diagrams contributing to
muon decay
we have
\be
2 \dg = \Vwu{1} +\Bwu{1} - \dy \,, \label{e27}
\ee
where the superscript $(1)$ reminds us that we are considering
one-loop
contributions. In \equ{e27} \Vwu{1}\ and \Bwu{1}\ represent the
vertex and box
diagrams. Their expressions can be gleaned from section 4 of
\efe{b12}.
Explicitly
\ben \label{e28} \beq
\Vwu{1} &= \frac{g^2}{16\pi^2}\left(\frac{4}{\epsilon}+4{\cal C} -4
\log\frac{\mz^2}{\mu^2}\right) \label{e28a} \\
\Bwu{1} &=\frac{5}{2} \frac{g^2}{16\pi^2}. \label{e28b}
\eeq\een

Consider now the two-loop contributions to the process. Some of the
relevant diagrams are shown in fig.~5. The group of graphs
represented
in fig.~5a gives a contribution
\ben \label{e30}
\be
M_{5a} = \left( \dy \right)^2 M_0 \,, \label{e30a}
\ee
where $M_0$ is the zeroth-order amplitude. Analogously, we have for
fig.~5b
\be
M_{5b} = - \dy \Vwu{1} M_0 \,. \label{e30b}
\ee
We now discuss 1PI vertex and box diagrams. The only graphs that can
give
an
$\mt^2$ contribution are those where a fermionic \self\ is inserted
in a vector
boson propagator line (fig.~5c). It is easy to show, using arguments
similar
to the ones developed in section 4 of \efe{b16}, that the $O(\gmud)$
part
of these diagrams can be obtained by considering the \self\ insertion
with
momentum transfer set equal to zero. Therefore these diagrams, plus
the
corresponding ones with a mass counterterm for the vector bosons,
give
to $O(\gmud)$:
\beq
M_{5c} =& \left( \Vwu{2} + \Bwu{2} \right) \dy M_0 \label{e30c} \\
\Vwu{2} =& - 2 \frac{g^2}{16 \pi^2} \label{e30d}\\
\Bwu{2} =& - \frac54 \frac{g^2}{16 \pi^2}. \label{e30e}
\eeq
To complete the calculation of $2 \delta g / g$, we have to consider
the reducible diagrams. Figure~5d gives a contribution
\be
M_{5d} = \left(  \dg \right)^2 M_0 \,, \label{e30f}
\ee
while expanding the coupling constants in the one-loop diagrams of
fig.~4
one obtains
\be
M_{exp} = - 4 \dg \left( \Vwu{1} + \Bwu{1} - \dy \right) M_0
\,.\label{e30g}
\ee
\een
Putting together \eqs{e30}\ and taking into account the two-loop 1PI
diagrams
analogous to fig.~4a,b, we can write for $\delta g / g$ to the order
we
are interested in
\beq
\frac{2\delta g}{g}=&
\left.\frac{{\rm Re}\azz(\mz^2)-\aww(0)}{\mz^2}\right|_{1PI}
 + \Vwu{1}+\Bwu{1} \nonumber\\
& +\left(\dg \right)^2 + \left( 3\Vwu{1}+\Vwu{2}+ 4 \Bwu{1} +
\Bwu{2} - \dy \right) \dy \,, \label{e31}
\eeq
where the first term includes both one- and two-loop 1PI
contributions to
the \selfs.

Having defined our counterterms we are now ready to consider
neutral-current
processes; in particular we will focus on neutrino--electron
scattering. With
our choice of renormalization conditions the interaction strength of
the
charged current is set equal to $G_\mu / \sqrt{2}$, therefore the
neutral-current amplitude
divided by $G_\mu / \sqrt{2}$ will give us the $\rho$ parameter
directly.

We refer again to fig.~4 and fig.~5 for one- and two-loop diagrams,
in this
case relevant to neutrino-electron scattering. The one-loop
contributions
of fig.~4 give
\bea
\dro{1} &=&  \frac{{\rm Re} \Azz{1} (\mz^2)-
            \Azz{1} (0)}{\mz^2} + \Vzu{1} +\Bzu{1}
       - 2 \dg \nonumber \\
        &=&  \frac{\Aww{1}(0)-
            \Azz{1} (0)}{\mz^2} + \Vzu{1} - \Vwu{1}
          +\Bzu{1} - \Bwu{1} \,, \label{e32}
\eea
where $\Vwu{1}=\Vzu{1}$ and
\be
\Bzu{1} = \frac{g^2}{16 \pi^2} \left( \frac52 -\frac94 I_3 \right).
\label{e33}
\ee
In \equ{e33}\ we have explicitly shown the part of the box diagrams
that is
process-dependent, i.e.\ the contribution proportional to $I_3$, the
isospin
of the target ($I_3 =-1$ for electrons).

We now proceed to examine the neutral process at the two-loop level.
We can neglect
diagrams 5a and 5b because the combination ${\rm
Re}\azz(\mz^2)-\azz(0)$ does
not contain $\mt^2$ terms. The sum of all the other diagrams gives a
contribution to $\rho$ equal to
\beq
\dr =& \left. \frac{{\rm Re}\azz(\mz^2)-\azz(0)}{\mz^2} \right|_{1PI}
+ V_\smallz
      + B_\smallz +\left( \dg \right)^2 \nonumber \\
&      - 4 \dg \left( \frac{{\rm Re}\Azz{1}(\mz^2)-
       \Azz{1} (0)}{\mz^2} + \Vzu{1} +\Bzu{1} \right)
       - 2 { \delta g \over g }, \label{e34}
\eeq
where $V_\smallz = \Vzu{1} +\Vzu{2} \dy$, $B_\smallz = \Bzu{1}
+\Bzu{2}
\dy$,
and
\ben \label{e35} \beq
\Vzu{2} =& - 4 \frac{g^2}{16 \pi^2} \label{e35a}\\
\Bzu{2} =& -  \frac{g^2}{16 \pi^2} \left[ \frac52 - \frac32 I_3
\right] .
           \label{e35b}
\eeq
\een
Recalling that
\be
\dy \simeq N_c x_t \left[ 1 + \epsilon \left( \frac12 + {\cal C} -
\ln
      \frac{\mt^2}{\mu^2} \right) \right] \label{e36}
\ee
we can write, after some simple algebra, the one- and two-loop
contribution
to the $\rho$ parameter up to $O(\gmud)$ terms as
\ben \label{e37}
\beq
\dro{1} =& x_t \left( N_c - 9 I_3 zt \right) \label{e37a} \\
\dro{2} =& \left.
\frac{\Aww{2}(0)- \Azz{2}(0)}{\mz^2}\right|_{1PI} \nonumber\\
& + N_c x_t^2  \left[ N_c -4 zt \left( {4\over\epsilon}+
                  8{\cal C} - 4 \ln{\mz^2\over\mu^2}
-4\ln{m_t^2\over\mu^2}
                +{41\over4} + 3 I_3 \right) \right] \,. \label{e37b}
\eeq
\een
Substituting the sum of eqs.~(24) and (27) in the first term of
eq.~(\ref{e37b}) we get our final result. The $1/ \epsilon $ pole
present in the
first term of \equ{e37b}\ (cf.\ \equ{e23}) is cancelled by an
analogous
contribution coming from the vertex part.

In table 1 we present $\dro{2}$ for $zt=0,\,0.2,\,0.25,\,0.3$ and
for several values
of $ht$. The entries in the table report the case $N_c=3$ and
$I_3=-1$.
The column $zt=0$ coincides with the results obtained in
\efs{b9}{b11}.
In all entries, the difference by a factor of 3
between the numbers reported in our
table and the ones presented in \efs{b9}{b11} is due to the fact that
\equ{e37b}
contains also the leading reducible contribution, i.e.\ the term
$(N_c x_t)^2$,
while the tables in \efs{b9}{b11} present only the two-loop
irreducible part.
In order to avoid numerical problems in the region $ht \simeq zt$ we
have
prepared the table employing the exact expressions for the diagrams
involving
the Higgs. The remaining contributions have been included to $O(zt)$.
Using
the asymptotic expressions of \eqs{e20}, we get a difference with
respect to the
numbers of table 1 by, at most, 3\%.

As the table shows, the $O(zt)$ terms have a dramatic effect
especially for
light Higgs-mass values.
This is easily understood by looking at \equ{e37b}. In fact the
finite
part\footnote{The term proportional to ${\cal C}$ is not included in
the finite
              part.}  of the term proportional to $zt$ in the second
row of
\equ{e37b} amounts, with $\mu=\mt$, to $-10.9,\, -12.8,\, -14.5$ for
$zt =0.2,\,0.25,\,0.3$, by far the largest contribution to $\dro{2}$
for small
$ht$. As $ht$ increases, the contribution of the \selfs\ starts to be
more
important. It is interesting to consider the asymptotic expansion of
the finite
part of the two-loop $\mt$ contribution to $\Delta_2$ for large $ht$
\beqs
\Delta_2 \simeq & N_c x_t^2 \left\{
          \frac{49}4 + \pi^2 - \frac{27}2 \ln ht + \frac32 \ln^2 ht
          + zt \left( 46 - \frac73 \pi^2 - \ln zt \right) \right\} \\
&+ \frac1{ht} \left[ \frac23 - 4 \pi^2 - 4 \ln ht - 9 \ln^2 ht
    + zt \left( \frac32 - 9 \ln ht - 3 \ln^2 ht \right) \right] \\
& + \frac1{ht^2} \left[ \frac{1613}{48} - 5 \pi^2 + \frac{125}4 \ln
ht
    - 15 \ln^2 ht \right. \\
& \hphantom{\frac1{ht^2} \frac{1613}{48}}
+ \left. \left. zt \left( \frac{124}{9} - 4 \pi^2 + \frac{16}3 \ln ht
- 10 \ln^2 ht
  \right)  \right] + \dots  \right\}. \yesnumber \label{e40}
\eeqs
We note that in the limit $ht \rightarrow \infty$ the inclusion of
subleading
contributions to the \selfs\ gives rise to an $\mh$-independent term.

\section{Conclusions}
In the previous sections we have shown that the use of current
correlation
functions and their associated current algebra provides, in the
fermionic
sector, an efficient way to enforce the relevant Ward identities
while
discussing at the same time several Feynman diagrams. In particular,
we
have seen in section 2 that, for the leading term, the formalism is
equivalent to the effective Lagrangian approach of \efe{b9}, sharing
with it the same advantages from a computational point of view.
Although
with respect to \efe{b9} we had to pay the price to explicitly derive
the Ward identities, this effort was rewarded by gaining a formalism
that could deal at the same time with subleading terms.

The calculation of \dr\ to $O(\gmud)$ shows some interesting
features.
Differently from the one-loop case where $\Delta_2$ is finite
and gauge-invariant, at the two-loop level,
beyond the leading $\mt^4$ term,
neither of these two properties is kept and therefore $\Delta_2$
cannot
be identified with a physical observable. As a comparison,
a similar situation happens in the SM already at the one-loop level.
In
fact,
there, one-loop gauge-dependent bosonic contributions do not cancel
in
the difference of the two \selfs.

Two, related, consequences of the above fact are: {\it i)} to define
a physical
observable, we have to resort to physical processes and this
introduces
process-dependent quantities. \linebreak
{\it ii)} Concerning the $\mt^4$ contribution,
the two-loop reducible and irreducible parts are separately finite.
Furthermore, the reducible term is equal to $(N_c x_t)^2$, namely
the square of the one-loop contribution. This fact suggested to the
authors
of \efe{b17} that a possible way to take into account higher-order
effects is to write $\rho$ as
\be
\rho \equiv \frac1{1 -\dr_{irr}} \,, \label{e50}
\ee
where
\be
\dr_{irr} = \dr^{(1)}_{irr} + \dr^{(2)}_{irr}\,,  \label{e51}
\ee
with $\dr^{(1)}_{irr}$ and $\dr^{(2)}_{irr}$ the 1PI one- and
two-loop
contribution to $\Delta$. There is no actual justification why
\equ{e50}
should be the correct way to resum higher-order top effects, its
validity
relying upon
the fact that up to two-loops it reproduces the correct result. We
have
seen that once subleading contributions are included, $\Delta_2$,
the equivalent of $\dr^{(2)}_{irr}$ in the $SU(2)$ model,
is not  finite. We expect that if we were to evaluate
$\dr^{(2)}_{irr}$ in the
SM to subleading order a similar situation would prevail. It then
seems that
\equ{e50} cannot work at the subleading level.
The simplest choice of replacing $\dr^{(2)}_{irr}$
by $\dro{2}$ in \equ{e50} cannot be theoretically justified. In fact
the scattering amplitude contains vertex and box
contributions (cf.\ \equ{e37b}), and it is not yet
clear how this type of diagrams can be resummed. To avoid this
problem,
one can
consider to resum only the leading top contributions. However, at the
two-loop
level there are diagrams (cf.\ fig.~1) that contribute both to the
leading
and the subleading part. The choice to split a diagram contribution
into two parts, one of which is resummed, seems quite arbitrary.
Probably the ``safest'' way to resum
higher-order effects is to assume
\be
\rho =\frac1{1 -\dr_{irr}^{(1)}} + \dr_{irr}^{(2)} \,, \label{e52}
\ee
where the resummation of $\dr_{irr}^{(1)}$ can be theoretically
justified
on the basis of $1/ N_c$ expansion arguments \cite{b50}.
Although there is no practical implication in choosing \equ{e52}\
instead of \equ{e50}, the two expressions have a different behaviour
with
respect to the limit $\mt \rightarrow \infty$. While in \equ{e50}, as
$\mt$ grows the two-loop term starts to cancel a large part
of the one-loop contribution and, eventually, for reasonable values
of
$\mh$, overwhelms it, leaving us with an upper bound on the $\rho$
parameter,
the same does not happen in \equ{e52}. Indeed \equ{e52} does not
allow
for screening of heavy physics by higher-order effects and the
$\rho$ parameter, in this case, is not bounded from above.

Finally, we want to comment on the size of subleadings effects.
As shown in table~1, the leading $O(\gmuq)$ contribution and the
subleading $O(\gmud)$
one are numerically comparable for realistic values of the top mass
and,
moreover, the two contributions have the same sign. The effect of
$O(\gmuq)$ correction, for $\mt= 200$ GeV, has been estimated
\cite{b51}
to decrease the value of the predicted $W$ mass in the SM by $\sim
24$ MeV.
Assuming that subleading effects were comparable to the leading
contribution, the shift in the prediction of the $W$ mass would be
almost
as large as the envisaged experimental accuracy $(\delta \mw)_{exp}
= \pm 50$ MeV. However, it is not correct to directly apply this
result
to LEP physics for several reasons. First, the calculation
of the subleading effects
we presented was performed in an $SU(2)$ model, where electromagnetic
interactions are not present. Secondly, for what concerns LEP
physics,
it is useful to compute $\sin^2 \theta_W$ as accurately as possible.
The
relation between the standard input parameters, $G_\mu,\, \mz$,
$\alpha$, and $\sin^2 \theta_W$ involves a correction factor
($\Delta r$ in the on-shell scheme, $\Delta \hat{r}$ in the
$\overline{MS}$
scheme
\cite{b52}) that contains \selfs\ evaluated at momentum transfers
equal
to the mass of the vector bosons and not at $q^2=0$. This fact
introduces
additional subleading $O(\gmud)$ terms that are not clearly taken
into
account in this paper. Furthermore, the vertex and box parts entering
in $\Delta r$ or $\Delta \hat{r}$ are different from the ones present
in
\dr. Moreover the calculation of the reducible contributions in the
SM
is more complicated because two coupling constants, $g$ and
$g^\prime$,
have to be taken into account.

To conclude, it is fair to say that the subleading two-loop $\mt$
effects
can be larger than what is ``nai\"vely'' expected. Although it is
quite
unlikely that these effects can modify the theoretical predictions of
the
various observables by amounts larger than the foreseeable
experimental
accuracy, it is realistic to assume that their influence can be
comparable
to,
or maybe larger than, the uncertainty in the evaluation of the
hadronic
contribution to the photonic \self.

\section*{Acknowledgements}
The authors want to thank F.~Feruglio, A.~Masiero, S.~Peris,
M.~Porrati,
A.~Santamaria
and A.~Sirlin for valuable discussions.
In particular we thank S.~Peris for bringing to our attention the
work of
\efe{b11bis}.
Two of us (G.D. and S.F.) would like
to thank the Physics Department of New York University for its kind
hospitality during November 1993, when part of this work was carried
out.
This research was supported in part by the National Science
Foundation under
Grant No.\ PHY-9017585. One of 	us (S.F.) is currently supported by
an
ICSC World Laboratory Fellowship at the CERN Theory Division.

\newpage
\section*{Figure captions}
\begin{description}
\item[Fig.~1:] Two-loop diagrams contributing to  $O(\gmuq)$  to
             the vector boson propagators. In the figure wavy lines
             represent vector bosons, dashed lines indicate scalars
             while solid lines are fermions.
\item[Fig.~2:] Relevant two-loop diagrams contributing to $\Pi_{\Phi
\Phi}$
             and $\Pi_{\Phi_2 \Phi_2}$.
\item[Fig.~3:] Some of the two-loop diagrams involving the Higgs
boson
	     contributing
             to $O(\gmud)$ to the vector boson propagators.
\item[Fig.~4:] One-loop graphs relevant to charged- or
neutral-current
             interactions. The shadowed blob in figs.~4c and 4d
indicates
             schematically the sum of graphs in which the virtual
gauge
             bosons are attached in all possible ways to the external
lines.
\item[Fig.~5:] Two-loop graphs relevant to charged- or
neutral-current
             interactions. The meaning of the shadowed blob is as in
fig.~4.
\end{description}
\clearpage
\newpage
\vfil
\par\centerline{\psfig{figure=fig1.ps,height=10cm}}
\vspace{0.4 true cm}
\centerline{\bf Fig.~1}\par
\vfill
\par\centerline{\psfig{figure=fig2.ps,height=7cm}}
\vspace{0.4 true cm}
\centerline{\bf Fig.~2}\par
\vfil
\newpage
\vfil
\par\centerline{\psfig{figure=fig3.ps,height=9cm}}
\vspace{0.4 true cm}
\centerline{\bf Fig.~3}\par
\vfill
\par\centerline{\psfig{figure=fig4.ps,height=7.5cm}}
\vspace{0.4 true cm}
\centerline{\bf Fig.~4}\par
\vfil
\newpage
\par\centerline{\psfig{figure=fig5.ps,height=19cm}}
\vspace{0.4 true cm}
\centerline{\bf Fig.~5}\par

\renewcommand{\arraystretch}{1.2}
\begin{table}[h]
\centering
\caption[TableI]{Values of $\delta \rho^{(2)}$, for $N_c=3,\,I_3=-1$
and
                 in units $N_c\,x_t^2$, as a function of $\mh /\mt$
                 for some values of $zt \equiv \mz^2 /\mt^2$}

\begin{tabular}{|c|c|c|c|c|}
\hline
$\mh/\mt$ &  \multicolumn{4}{c|}{ $ \delta\rho^{(2)}$ } \\
\cline{2-5}
  {} & $zt = 0$ & \phantom{$zt$}0.2\phantom{$zt$}  &
  \phantom{$zt$}0.25\phantom{$zt$}  & \phantom{$zt$}0.3\phantom{$zt$}
\\
        \hline
  0.10    & \phantom{$-$}1.18   &    $-$8.28    &   $-$9.56    &
$-$10.7 \\
  0.20    & \phantom{$-$}0.30   &    $-$8.96    &   $-$10.2    &
$-$11.3 \\
  0.30    &         $-$0.46  &    $-$9.51    &   $-$10.8    &
$-$11.8 \\
  0.40    &         $-$1.13  &    $-$10.0    &   $-$11.2    &
$-$12.2 \\
  0.50    &         $-$1.72  &    $-$10.4    &   $-$11.6    &
$-$12.6 \\
  0.60    &         $-$2.25  &    $-$10.7    &   $-$11.9    &
$-$12.9 \\
  0.70    &         $-$2.74  &    $-$11.1    &   $-$12.2    &
$-$13.2 \\
  0.80    &         $-$3.18  &    $-$11.4    &   $-$12.5    &
$-$13.4 \\
  0.90    &         $-$3.58  &    $-$11.6    &   $-$12.7    &
$-$13.7 \\
  1.00    &         $-$3.96  &    $-$11.9    &   $-$13.0    &
$-$13.9 \\
  1.10    &         $-$4.30  &    $-$12.1    &   $-$13.2    &
$-$14.1 \\
  1.20    &         $-$4.62  &    $-$12.4    &   $-$13.4    &
$-$14.3 \\
  1.30    &         $-$4.91  &    $-$12.6    &   $-$13.6    &
$-$14.5 \\
  1.40    &         $-$5.19  &    $-$12.8    &   $-$13.8    &
$-$14.6 \\
  1.50    &         $-$5.44  &    $-$12.9    &   $-$13.9    &
$-$14.8 \\
  1.60    &         $-$5.68  &    $-$13.1    &   $-$14.1    &
$-$14.9 \\
  1.70    &         $-$5.90  &    $-$13.3    &   $-$14.2    &
$-$15.1 \\
  1.80    &         $-$6.11  &    $-$13.4    &   $-$14.4    &
$-$15.2 \\
  1.90    &         $-$6.30  &    $-$13.6    &   $-$14.5    &
$-$15.3 \\
  2.00    &         $-$6.48  &    $-$13.7    &   $-$14.6    &
$-$15.4 \\
  2.50    &         $-$7.24  &    $-$14.2    &   $-$15.1    &
$-$15.9 \\
  3.00    &         $-$7.78  &    $-$14.6    &   $-$15.5    &
$-$16.2 \\
  3.50    &         $-$8.17  &    $-$14.9    &   $-$15.7    &
$-$16.4 \\
  4.00    &         $-$8.44  &    $-$15.1    &   $-$15.9    &
$-$16.6 \\
  4.50    &         $-$8.61  &    $-$15.2    &   $-$16.0    &
$-$16.6 \\
  5.00    &         $-$8.72  &    $-$15.2    &   $-$16.0    &
$-$16.6 \\
  5.50    &         $-$8.76  &    $-$15.2    &   $-$16.0    &
$-$16.6 \\
  6.00    &         $-$8.76  &    $-$15.2    &   $-$15.9    &
$-$16.6 \\
\hline
\end{tabular}
\end{table}
\end{document}